\PassOptionsToPackage{unicode}{hyperref}
\PassOptionsToPackage{hyphens}{url}
\PassOptionsToPackage{dvipsnames,svgnames,x11names}{xcolor}
\documentclass[
]{article}
\usepackage{amsmath,amssymb}
\usepackage{lmodern}
\usepackage{iftex}
\ifPDFTeX
  \usepackage[T1]{fontenc}
  \usepackage[utf8]{inputenc}
  \usepackage{textcomp} 
\else 
  \usepackage{unicode-math}
  \defaultfontfeatures{Scale=MatchLowercase}
  \defaultfontfeatures[\rmfamily]{Ligatures=TeX,Scale=1}
\fi
\IfFileExists{upquote.sty}{\usepackage{upquote}}{}
\IfFileExists{microtype.sty}{
  \usepackage[]{microtype}
  \UseMicrotypeSet[protrusion]{basicmath} 
}{}
\makeatletter
\@ifundefined{KOMAClassName}{
  \IfFileExists{parskip.sty}{%
    \usepackage{parskip}
  }{
    \setlength{\parindent}{0pt}
    \setlength{\parskip}{6pt plus 2pt minus 1pt}}
}{
  \KOMAoptions{parskip=half}}
\makeatother
\usepackage{xcolor}
\usepackage{graphicx}
\makeatletter
\def\maxwidth{\ifdim\Gin@nat@width>\linewidth\linewidth\else\Gin@nat@width\fi}
\def\maxheight{\ifdim\Gin@nat@height>\textheight\textheight\else\Gin@nat@height\fi}
\makeatother
\setkeys{Gin}{width=\maxwidth,height=\maxheight,keepaspectratio}
\makeatletter
\def\fps@figure{htbp}
\makeatother
\NewDocumentCommand\citeproctext{}{}
\NewDocumentCommand\citeproc{mm}{%
  \begingroup\def\citeproctext{#2}\cite{#1}\endgroup}
\makeatletter
 \let\@cite@ofmt\@firstofone
 \def\@biblabel#1{}
 \def\@cite#1#2{{#1\if@tempswa , #2\fi}}
\makeatother
\newlength{\cslhangindent}
\newlength{\csllabelwidth}
\newenvironment{CSLReferences}[2] 
 {\begin{list}{}{%
  \setlength{\itemindent}{0pt}
  \setlength{\leftmargin}{0pt}
  \setlength{\parsep}{0pt}
  \ifodd #1
   \setlength{\leftmargin}{\cslhangindent}
   \setlength{\itemindent}{-1\cslhangindent}
  \fi
  \setlength{\itemsep}{#2\baselineskip}}}
 {\end{list}}
\usepackage{calc}

\ifLuaTeX
\usepackage[bidi=basic]{babel}
\else
\usepackage[bidi=default]{babel}
\fi
\babelprovide[main,import]{american}

\def\languageshorthands#1{}
\ifLuaTeX
  \usepackage{selnolig}  
\fi
\IfFileExists{bookmark.sty}{\usepackage{bookmark}}{\usepackage{hyperref}}
\IfFileExists{xurl.sty}{\usepackage{xurl}}{} 
\hypersetup{
  pdftitle={Million Points of Light (MPoL): a PyTorch library for radio
interferometric imaging and inference},
  pdfauthor={Ian Czekala, Jeff Jennings, Brianna Zawadzki, Kadri Nizam,
Ryan Loomis, Megan Delamer, Kaylee de Soto, Robert Frazier, Hannah
Grzybowski, Mary Ogborn, Tyler Quinn},
  pdflang={en-US},
  colorlinks=true,
  linkcolor={Maroon},
  filecolor={Maroon},
  citecolor={Blue},
  urlcolor={Blue},
  pdfcreator={LaTeX via pandoc}}

\title{Million Points of Light (MPoL): a PyTorch library for radio
interferometric imaging and inference}

\definecolor{c53baa1}{RGB}{83,186,161}
\definecolor{c202826}{RGB}{32,40,38}

\usepackage[affil-it]{authblk}
\usepackage{orcidlink}
\author[1%
  \ensuremath\mathparagraph]{Ian Czekala%
    \,\orcidlink{0000-0002-1483-8811}\,%
    }
\author[2%
  ]{Jeff Jennings%
    \,\orcidlink{0000-0002-7032-2350}\,%
    }
\author[3%
  ]{Brianna Zawadzki%
    \,\orcidlink{0000-0001-9319-1296}\,%
    }
\author[4%
  ]{Kadri Nizam%
    \,\orcidlink{0000-0002-7217-446X}\,%
    }
\author[5%
  ]{Ryan Loomis%
    \,\orcidlink{0000-0002-8932-1219}\,%
    }
\author[4%
  ]{Megan Delamer%
    \,\orcidlink{0000-0003-1439-2781}\,%
    }
\author[4%
  ]{Kaylee de Soto%
    \,\orcidlink{0000-0002-9886-2834}\,%
    }
\author[4%
  ]{Robert Frazier%
    \,\orcidlink{0000-0001-6569-3731}\,%
    }
\author[4%
  ]{Hannah Grzybowski%
    }
\author[6]{Jane Huang\,\orcidlink{0000-0001-6947-6072}\,}
\author[4%
  ]{Mary Ogborn%
    \,\orcidlink{0000-0001-9741-2703}\,%
    }
\author[4%
  ]{Tyler Quinn%
    \,\orcidlink{0000-0002-8974-8095}\,%
    }

\affil[1]{University of St Andrews, Scotland%
  }
\affil[2]{CCA, Flatiron Institute, NY, USA%
  }
\affil[3]{Wesleyan University, CT, USA%
  }
\affil[4]{Pennsylvania State University, PA, USA%
  }
\affil[5]{National Radio Astronomy Observatory, Charlottesville, VA,
USA
  }
\affil[6]{Columbia University, NY, USA
  }
\affil[$\mathparagraph$]{Corresponding author: %
}
\date{24 January 2025}

\begin{document}
\maketitle

\section{Summary}\label{summary}

Astronomical radio interferometers achieve exquisite angular resolution
by cross-correlating signal from a cosmic source simultaneously observed
by distant pairs of radio telescopes to produce a Fourier-type
measurement called a visibility. \emph{Million Points of Light}
(\texttt{MPoL}) is a Python library supporting feed-forward modeling of
interferometric visibility datasets for synthesis imaging and parametric
Bayesian inference, built using the autodifferentiable machine learning
framework PyTorch. Neural network components provide a rich set of
modular and composable building blocks that can be used to express the
physical relationships between latent model parameters and observed data
following the radio interferometric measurement equation. Industry-grade
optimizers make it straightforward to simultaneously solve for the
synthesized image and calibration parameters using stochastic gradient
descent.

\section{Statement of need}\label{statement-of-need}

When an astrophysical source is observed by a radio interferometer,
there are frequently large gaps in the spatial frequency coverage.
Therefore, rather than perform a direct Fourier inversion, images must
be synthesized from the visibility data using an imaging algorithm; it
is common for the incomplete sampling to severely hamper image fidelity
(\citeproc{ref-condon16}{Condon \& Ransom, 2016};
\citeproc{ref-thompson17}{Thompson et al., 2017}). CLEAN is the
traditional image synthesis algorithm of the radio interferometry
community (\citeproc{ref-hogbom74}{Högbom, 1974}), with a modern
implementation in the reduction and analysis software CASA
(\citeproc{ref-casa22}{CASA Team et al., 2022};
\citeproc{ref-mcmullin07}{McMullin et al., 2007}), the standard for
current major facility operations (e.g., \citeproc{ref-hunter23}{Hunter
et al., 2023}). CLEAN excels at the rapid imaging of astronomical fields
comprising unresolved point sources (e.g.~quasars) and marginally
resolved sources, but may struggle when the source morphology is not
well-matched by the CLEAN basis set (e.g., point sources, Gaussians), a
common situation with ring-like protoplanetary disk sources
(\citeproc{ref-disk20}{Disk Dynamics Collaboration et al., 2020, sec.
3}).

High fidelity imaging algorithms for spatially resolved sources are
needed to realize the full scientific potential of groundbreaking
observatories like the Atacama Large Millimeter Array (ALMA; Wootten \&
Thompson (\citeproc{ref-wootten09}{2009})), the Event Horizon Telescope
(\citeproc{ref-eht19a}{Event Horizon Telescope Collaboration, et
al., 2019}), and the Square Kilometer Array
(\citeproc{ref-dewdney09}{Dewdney et al., 2009}) as they deliver
significantly improved sensitivity and resolving power compared to
previous generation instruments. In the field of planet formation alone,
spatially resolved observations from ALMA have rapidly advanced our
understanding of protoplanetary disk structures
(\citeproc{ref-andrews20}{Andrews, 2020}), kinematic signatures of
embedded protoplanets (\citeproc{ref-pinte18}{Pinte et al., 2018}), and
circumplanetary disks (\citeproc{ref-benisty21}{Benisty et al., 2021};
\citeproc{ref-casassus22}{Casassus \& Cárcamo, 2022}). Application of
higher performance imaging techniques to these groundbreaking datasets
(e.g., \citeproc{ref-casassus22}{Casassus \& Cárcamo, 2022}) showed
great promise in unlocking further scientific progress. Simultaneously,
a flexible, open-source platform could integrate machine learning
algorithms and computational imaging techniques from non-astronomy
fields.

\section{The Million Points of Light (MPoL)
library}\label{the-million-points-of-light-mpol-library}

\texttt{MPoL} is a library designed for feed-forward modeling of
interferometric datasets using Python, Numpy
(\citeproc{ref-harris20}{Harris et al., 2020}), and the computationally
performant machine learning framework PyTorch
(\citeproc{ref-paszke19}{Paszke et al., 2019}), which debuted with
Zawadzki et al. (\citeproc{ref-zawadzki23}{2023}). \texttt{MPoL}
implements a set of foundational interferometry components using PyTorch
\texttt{nn.module}, which can be easily combined to build a
forward-model of the interferometric dataset(s) at hand. We strive to
seamlessly integrate with the PyTorch ecosystem so that users can easily
leverage well-established machine learning workflows: optimization with
stochastic gradient descent (\citeproc{ref-bishop23}{Bishop \& Bishop,
2023}, Ch. 7), straightforward acceleration with GPU(s), and integration
with common neural network architectures.

In a typical feed-forward workflow, \texttt{MPoL} users will use
foundational components like \texttt{BaseCube} and \texttt{ImageCube} to
define the true-sky model, Fourier layers like \texttt{FourierCube} or
\texttt{NuFFT} (wrapping \texttt{torchkbnufft},
\citeproc{ref-nufft20}{Muckley et al., 2020}) to apply the Fourier
transform and sample the visibility function at the location of the
array baselines, and the negative log likelihood to calculate a data
loss. Backpropagation (see \citeproc{ref-baydin18}{Baydin et al., 2018}
for a review) and stochastic gradient descent (e.g., AdamW,
\citeproc{ref-loshchilov17}{Loshchilov \& Hutter, 2017}) are used to
find the true-sky model that minimizes the loss function. However,
because of the aforementioned gaps in spatial frequency coverage, there
is technically an infinite number of true-sky images fully consistent
with the data likelihood, so regularization loss terms are required.
\texttt{MPoL} supports Regularized Maximum Likelihood (RML) imaging with
common regularizers like maximum entropy, sparsity, and others (e.g., as
used in \citeproc{ref-eht19d}{Event Horizon Telescope Collaboration 2019}); users can also implement custom regularizers with
PyTorch.

\texttt{MPoL} also provides several other workflows relevant to
astrophysical research. First, by seamlessly coupling with the
probabilistic programming language Pyro (\citeproc{ref-pyro19}{Bingham
et al., 2019}), \texttt{MPoL} supports Bayesian parametric inference of
astronomical sources by modeling the data visibilities. Second, users
can implement additional data calibration components as their data
requires, enabling fine-scale, residual calibration physics to be
parameterized and optimized simultaneously with image synthesis
(following the radio interferometric measurement equation
\citeproc{ref-hamaker96}{Hamaker et al., 1996};
\citeproc{ref-smirnov11a}{Smirnov, 2011}). Finally, the library also
provides convenience utilities like \texttt{DirtyImager} (including
Briggs robust and UV taper) to confirm the data has been loaded
correctly. The MPoL-dev organization also develops the
\href{https://mpol-dev.github.io/visread/}{MPoL-dev/visread} package,
which is designed to facilitate the extraction of visibility data from
CASA's Measurement Set format for use in alternative imaging workflows.

\section{Documentation, examples, and scientific
results}\label{documentation-examples-and-scientific-results}

\begin{figure}
\centering
\includegraphics{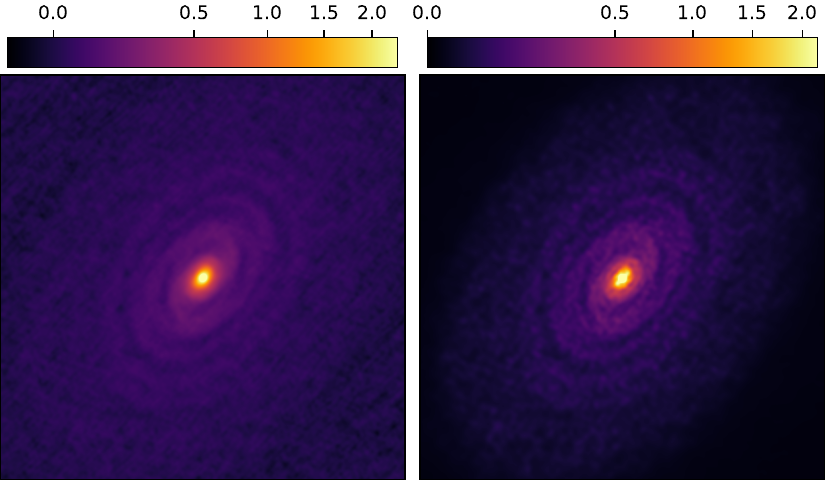}
\caption{Left: the synthesized image produced by the DSHARP ALMA Large
Program (\citeproc{ref-andrews18}{Andrews et al., 2018}) using
\texttt{CASA/tclean}. Right: The regularized maximum likelihood image
produced using \texttt{MPoL} on the same data. Both images are displayed
using a \texttt{sqrt} stretch, with upper limit truncated to 70\% and
40\% of max value for CLEAN and \texttt{MPoL}, respectively, to
emphasize faint features. The CLEAN algorithm permits negative intensity
values, while the \texttt{MPoL} algorithm enforces image positivity by
construction. Each side of the image is 3 arcseconds. Intensity units
are shown in units of Jy/arcsec\textsuperscript{2}. \label{imlup}}
\end{figure}

MPoL is freely available, open-source software licensed via the MIT
license and is developed on GitHub at
\href{https://github.com/MPoL-dev/MPoL}{MPoL-dev/MPoL}. Installation and
API documentation is hosted at \url{https://mpol-dev.github.io/MPoL/},
and is continuously built with each commit to the \texttt{main} branch.
As a library, \texttt{MPoL} expects researchers to write short scripts
using use \texttt{MPoL} and PyTorch primitives, in much the same way
that PyTorch users write scripts for machine learning workflows (e.g.,
as in the \href{https://github.com/pytorch/examples}{official PyTorch
examples}). \texttt{MPoL} example projects are hosted on GitHub at
\href{https://github.com/MPoL-dev/examples}{MPoL-dev/examples}. These
include an introduction to generating mock data, a quickstart using
stochastic gradient descent, and a Pyro workflow using stochastic
variational inference (SVI) to replicate the parametric inference done
in Guzmán et al. (\citeproc{ref-guzman18}{2018}), among others. In
Figure \ref{imlup}, we compare an image obtained with CLEAN to that
using \texttt{MPoL} and RML, synthesized from the data presented in
Huang et al. (\citeproc{ref-huang18b}{2018}), highlighting the
improvement in resolution offered by feed-forward modeling
technologies.\footnote{Source code to reproduce this result is available
  as an \href{https://github.com/MPoL-dev/examples/tree/main}{MPoL
  example}.}

\texttt{MPoL} has already been used in a number of scientific
publications. Zawadzki et al. (\citeproc{ref-zawadzki23}{2023})
introduced \texttt{MPoL} and explored RML imaging for ALMA observations
of protoplanetary disks, finding a 3x improvement in spatial resolution
at comparable sensitivity. Dia et al. (\citeproc{ref-dia23}{2023}) used
\texttt{MPoL} as a reference imaging implementation to evaluate the
performance of their score-based prior algorithm. Huang et al.
(\citeproc{ref-huang24}{2024}) used the parametric inference
capabilities of \texttt{MPoL} to analyze radial dust substructures in a
suite of eight protoplanetary disks in the \(\sigma\) Orionis stellar
cluster. \texttt{MPoL} was selected as an imaging technology of the
exoALMA large program, where Zawadzki et al.~2024 \emph{submitted} used
RML imaging to obtain high resolution image cubes of molecular line
emission in protoplanetary disks in order to identify non-Keplerian
features that may trace planet-disk interactions.

\section{Similar tools}\label{similar-tools}

Recently, there has been significant work to design robust algorithms to
image spatially resolved sources. A non-exhaustive list includes the
\texttt{RESOLVE} family of algorithms
(\citeproc{ref-junklewitz16}{Junklewitz et al., 2016}), which impose
Gaussian random field image priors, the multi-algorithm approach of the
Event Horizon Telescope Collaboration (\citeproc{ref-eht19d}{Event
Horizon Telescope Collaboration 2019}) including regularized
maximum likelihood techniques, MaxEnt (\citeproc{ref-carcamo18}{Cárcamo
et al., 2018}), and domain-specific non-parametric 1D approaches like
\texttt{frank} (\citeproc{ref-jennings20}{Jennings et al., 2020}).
Several approaches have leveraged deep-learning, such as score-based
priors (\citeproc{ref-dia23}{Dia et al., 2023}), denoising diffusion
probabilistic models (\citeproc{ref-wang23}{Wang et al., 2023}), and
residual-to-residual deep neural networks
(\citeproc{ref-dabbech24}{Dabbech et al., 2024}). By contrast to many
imaging software programs, \texttt{MPoL} is designed as a library, and
so in theory can support a variety of forward-modeling workflows.

The parametric modeling capabilities of \texttt{MPoL}, provided by
integration with \texttt{Pyro}, are similar to the \texttt{emcee}
(\citeproc{ref-foreman-mackey13}{Foreman-Mackey et al., 2013}) +
synthetic visibility workflow provided by the Galario software
(\citeproc{ref-tazzari18}{Tazzari et al., 2018}). Since PyTorch enables
automatic differentiation, \texttt{Pyro} users can utilize HMC/NUTS
sampling (\citeproc{ref-hoffman14}{Hoffman et al., 2014};
\citeproc{ref-neal12}{Neal, 2012}) or SVI, which offer significant
benefits in high dimensional spaces compared to ensemble MCMC samplers.

\section{Acknowledgements}\label{acknowledgements}

We acknowledge funding from an ALMA Development Cycle 8 grant number
AST-1519126. J.H. acknowledges support by the National Science
Foundation under Grant No.~2307916. ALMA is a partnership of ESO
(representing its member states), NSF (USA) and NINS (Japan), together
with NRC (Canada), MOST and ASIAA (Taiwan), and KASI (Republic of
Korea), in cooperation with the Republic of Chile. The Joint ALMA
Observatory is operated by ESO, AUI/NRAO and NAOJ. The National Radio
Astronomy Observatory is a facility of the National Science Foundation
operated under cooperative agreement by Associated Universities, Inc.

\section*{References}\label{references}
\addcontentsline{toc}{section}{References}

\phantomsection\label{refs}


\begin{CSLReferences}{1}{0}
\bibitem[\citeproctext]{ref-andrews20}
Andrews, S. M. (2020). {Observations of Protoplanetary Disk Structures}.
\emph{Annual Review of Astronomy and Astrophysics}, \emph{58}, 483--528.
\url{https://doi.org/10.1146/annurev-astro-031220-010302}

\bibitem[\citeproctext]{ref-andrews18}
Andrews, S. M., Huang, J., Pérez, L. M., Isella, A., Dullemond, C. P.,
Kurtovic, N. T., Guzmán, V. V., Carpenter, J. M., Wilner, D. J., Zhang,
S., Zhu, Z., Birnstiel, T., Bai, X.-N., Benisty, M., Hughes, A. M.,
Öberg, K. I., \& Ricci, L. (2018). {The Disk Substructures at High
Angular Resolution Project (DSHARP). I. Motivation, Sample, Calibration,
and Overview}. \emph{Astrophysical Journal Letters}, \emph{869}(2), L41.
\url{https://doi.org/10.3847/2041-8213/aaf741}

\bibitem[\citeproctext]{ref-baydin18}
Baydin, A. G., Pearlmutter, B. A., Radul, A. A., \& Siskind, J. M.
(2018). Automatic differentiation in machine learning: A survey.
\emph{Journal of Machine Learning Research}, \emph{18}(153), 1--43.
\url{http://jmlr.org/papers/v18/17-468.html}

\bibitem[\citeproctext]{ref-benisty21}
Benisty, M., Bae, J., Facchini, S., Keppler, M., Teague, R., Isella, A.,
Kurtovic, N. T., Pérez, L. M., Sierra, A., Andrews, S. M., Carpenter,
J., Czekala, I., Dominik, C., Henning, T., Menard, F., Pinilla, P., \&
Zurlo, A. (2021). {A Circumplanetary Disk around PDS70c}. \emph{The
Astrophysical Journal Letters}, \emph{916}(1), L2.
\url{https://doi.org/10.3847/2041-8213/ac0f83}

\bibitem[\citeproctext]{ref-pyro19}
Bingham, E., Chen, J. P., Jankowiak, M., Obermeyer, F., Pradhan, N.,
Karaletsos, T., Singh, R., Szerlip, P., Horsfall, P., \& Goodman, N. D.
(2019). Pyro: Deep universal probabilistic programming. \emph{J. Mach.
Learn. Res.}, \emph{20}(1), 973--978.

\bibitem[\citeproctext]{ref-bishop23}
Bishop, C. M., \& Bishop, H. (2023). \emph{Deep learning - foundations
and concepts} (S. Cham, Ed.; 1st ed.).
https://doi.org/\url{https://doi.org/10.1007/978-3-031-45468-4}

\bibitem[\citeproctext]{ref-carcamo18}
Cárcamo, M., Román, P. E., Casassus, S., Moral, V., \& Rannou, F. R.
(2018). Multi-GPU maximum entropy image synthesis for radio astronomy.
\emph{Astronomy and Computing}, \emph{22}, 16--27.
https://doi.org/\url{https://doi.org/10.1016/j.ascom.2017.11.003}

\bibitem[\citeproctext]{ref-casa22}
CASA Team, Bean, B., Bhatnagar, S., Castro, S., Donovan Meyer, J.,
Emonts, B., Garcia, E., Garwood, R., Golap, K., Gonzalez Villalba, J.,
Harris, P., Hayashi, Y., Hoskins, J., Hsieh, M., Jagannathan, P.,
Kawasaki, W., Keimpema, A., Kettenis, M., Lopez, J., \ldots{} Kern, J.
(2022). {CASA, the Common Astronomy Software Applications for Radio
Astronomy}. \emph{Publications of the Astronomical Society of the
Pacific}, \emph{134}(1041), 114501.
\url{https://doi.org/10.1088/1538-3873/ac9642}

\bibitem[\citeproctext]{ref-casassus22}
Casassus, S., \& Cárcamo, M. (2022). {Variable structure in the PDS 70
disc and uncertainties in radio-interferometric image restoration}.
\emph{513}(4), 5790--5798. \url{https://doi.org/10.1093/mnras/stac1285}

\bibitem[\citeproctext]{ref-condon16}
Condon, J. J., \& Ransom, S. M. (2016). \emph{{Essential Radio
Astronomy}}.

\bibitem[\citeproctext]{ref-dabbech24}
Dabbech, A., Aghabiglou, A., Chu, C. S., \& Wiaux, Y. (2024). {CLEANing
Cygnus A Deep and Fast with R2D2}. \emph{The Astrophysical Journal
Letters}, \emph{966}(2), L34.
\url{https://doi.org/10.3847/2041-8213/ad41df}

\bibitem[\citeproctext]{ref-dewdney09}
Dewdney, P. E., Hall, P. J., Schilizzi, R. T., \& Lazio, T. J. L. W.
(2009). {The Square Kilometre Array}. \emph{IEEE Proceedings},
\emph{97}(8), 1482--1496.
\url{https://doi.org/10.1109/JPROC.2009.2021005}

\bibitem[\citeproctext]{ref-dia23}
Dia, N., Yantovski-Barth, M. J., Adam, A., Bowles, M., Lemos, P.,
Scaife, A. M. M., Hezaveh, Y., \& Perreault-Levasseur, L. (2023).
{Bayesian Imaging for Radio Interferometry with Score-Based Priors}.
\emph{arXiv e-Prints}, arXiv:2311.18012.
\url{https://doi.org/10.48550/arXiv.2311.18012}

\bibitem[\citeproctext]{ref-disk20}
Disk Dynamics Collaboration, Armitage, P. J., Bae, J., Benisty, M.,
Bergin, E. A., Casassus, S., Czekala, I., Facchini, S., Fung, J., Hall,
C., Ilee, J. D., Keppler, M., Kuznetsova, A., Le Gal, R., Loomis, R. A.,
Lyra, W., Manger, N., Perez, S., Pinte, C., \ldots{} Zhang, K. (2020).
{Visualizing the Kinematics of Planet Formation}. \emph{arXiv e-Prints},
arXiv:2009.04345. \url{https://doi.org/10.48550/arXiv.2009.04345}

\bibitem[\citeproctext]{ref-eht19d}
Event Horizon Telescope Collaboration. \ldots{} (2019). {First M87 Event Horizon
Telescope Results. IV. Imaging the Central Supermassive Black Hole}.
\emph{The Astrophysical Journal Letters}, \emph{875}(1), L4.
\url{https://doi.org/10.3847/2041-8213/ab0e85}

\bibitem[\citeproctext]{ref-eht19a}
Event Horizon Telescope Collaboration. \ldots{}. (2019). {First M87 Event Horizon
Telescope Results. I. The Shadow of the Supermassive Black Hole}.
\emph{The Astrophysical Journal Letters}, \emph{875}(1), L1.
\url{https://doi.org/10.3847/2041-8213/ab0ec7}

\bibitem[\citeproctext]{ref-foreman-mackey13}
Foreman-Mackey, D., Hogg, D. W., Lang, D., \& Goodman, J. (2013).
{emcee: The MCMC Hammer}. \emph{125}(925), 306.
\url{https://doi.org/10.1086/670067}

\bibitem[\citeproctext]{ref-guzman18}
Guzmán, V. V., Huang, J., Andrews, S. M., Isella, A., Pérez, L. M.,
Carpenter, J. M., Dullemond, C. P., Ricci, L., Birnstiel, T., Zhang, S.,
Zhu, Z., Bai, X.-N., Benisty, M., Öberg, K. I., \& Wilner, D. J. (2018).
{The Disk Substructures at High Angular Resolution Program (DSHARP).
VIII. The Rich Ringed Substructures in the AS 209 Disk}. \emph{869}(2),
L48. \url{https://doi.org/10.3847/2041-8213/aaedae}

\bibitem[\citeproctext]{ref-hamaker96}
Hamaker, J. P., Bregman, J. D., \& Sault, R. J. (1996). {Understanding
radio polarimetry. I. Mathematical foundations.} \emph{117}, 137--147.

\bibitem[\citeproctext]{ref-harris20}
Harris, C. R., Millman, K. J., Walt, S. J. van der, Gommers, R.,
Virtanen, P., Cournapeau, D., Wieser, E., Taylor, J., Berg, S., Smith,
N. J., Kern, R., Picus, M., Hoyer, S., Kerkwijk, M. H. van, Brett, M.,
Haldane, A., Río, J. F. del, Wiebe, M., Peterson, P., \ldots{} Oliphant,
T. E. (2020). Array programming with {NumPy}. \emph{Nature},
\emph{585}(7825), 357--362.
\url{https://doi.org/10.1038/s41586-020-2649-2}

\bibitem[\citeproctext]{ref-hoffman14}
Hoffman, M. D., Gelman, A., \& others. (2014). The no-u-turn sampler:
Adaptively setting path lengths in hamiltonian monte carlo. \emph{J.
Mach. Learn. Res.}, \emph{15}(1), 1593--1623.

\bibitem[\citeproctext]{ref-hogbom74}
Högbom, J. A. (1974). {Aperture Synthesis with a Non-Regular
Distribution of Interferometer Baselines}. \emph{Astronomy and
Astrophysics Supplement}, \emph{15}, 417.

\bibitem[\citeproctext]{ref-huang18b}
Huang, J., Andrews, S. M., Pérez, L. M., Zhu, Z., Dullemond, C. P.,
Isella, A., Benisty, M., Bai, X.-N., Birnstiel, T., Carpenter, J. M.,
Guzmán, V. V., Hughes, A. M., Öberg, K. I., Ricci, L., Wilner, D. J., \&
Zhang, S. (2018). {The Disk Substructures at High Angular Resolution
Project (DSHARP). III. Spiral Structures in the Millimeter Continuum of
the Elias 27, IM Lup, and WaOph 6 Disks}. \emph{869}(2), L43.
\url{https://doi.org/10.3847/2041-8213/aaf7a0}

\bibitem[\citeproctext]{ref-huang24}
Huang, J., Ansdell, M., Birnstiel, T., Czekala, I., Long, F., Williams,
J., Zhang, S., \& Zhu, Z. (2024). {High-resolution ALMA Observations of
Richly Structured Protoplanetary Disks in {\(\sigma\)} Orionis}.
\emph{The Astrophysical Journal}, \emph{976}(1), 132.
\url{https://doi.org/10.3847/1538-4357/ad84df}

\bibitem[\citeproctext]{ref-hunter23}
Hunter, T. R., Indebetouw, R., Brogan, C. L., Berry, K., Chang, C.-S.,
Francke, H., Geers, V. C., Gómez, L., Hibbard, J. E., Humphreys, E. M.,
Kent, B. R., Kepley, A. A., Kunneriath, D., Lipnicky, A., Loomis, R. A.,
Mason, B. S., Masters, J. S., Maud, L. T., Muders, D., \ldots{} Yoon, I.
(2023). {The ALMA Interferometric Pipeline Heuristics}.
\emph{135}(1049), 074501. \url{https://doi.org/10.1088/1538-3873/ace216}

\bibitem[\citeproctext]{ref-jennings20}
Jennings, J., Booth, R. A., Tazzari, M., Rosotti, G. P., \& Clarke, C.
J. (2020). {frankenstein: protoplanetary disc brightness profile
reconstruction at sub-beam resolution with a rapid Gaussian process}.
\emph{Monthly Notices of the RAS}, \emph{495}(3), 3209--3232.
\url{https://doi.org/10.1093/mnras/staa1365}

\bibitem[\citeproctext]{ref-junklewitz16}
Junklewitz, H., Bell, M. R., Selig, M., \& Enßlin, T. A. (2016).
{RESOLVE: A new algorithm for aperture synthesis imaging of extended
emission in radio astronomy}. \emph{586}, A76.
\url{https://doi.org/10.1051/0004-6361/201323094}

\bibitem[\citeproctext]{ref-loshchilov17}
Loshchilov, I., \& Hutter, F. (2017). {Decoupled Weight Decay
Regularization}. \emph{arXiv e-Prints}, arXiv:1711.05101.
\url{https://doi.org/10.48550/arXiv.1711.05101}

\bibitem[\citeproctext]{ref-mcmullin07}
McMullin, J. P., Waters, B., Schiebel, D., Young, W., \& Golap, K.
(2007). {CASA Architecture and Applications}. In R. A. Shaw, F. Hill, \&
D. J. Bell (Eds.), \emph{Astronomical data analysis software and systems
XVI ASP conference series} (Vol. 376, p. 127).

\bibitem[\citeproctext]{ref-nufft20}
Muckley, M. J., Stern, R., Murrell, T., \& Knoll, F. (2020).
\emph{{TorchKbNufft}: A high-level, hardware-agnostic non-uniform fast
{Fourier} transform}.

\bibitem[\citeproctext]{ref-neal12}
Neal, R. M. (2012). MCMC using hamiltonian dynamics. \emph{arXiv
Preprint arXiv:1206.1901}.

\bibitem[\citeproctext]{ref-paszke19}
Paszke, A., Gross, S., Massa, F., Lerer, A., Bradbury, J., Chanan, G.,
Killeen, T., Lin, Z., Gimelshein, N., Antiga, L., Desmaison, A., Köpf,
A., Yang, E., DeVito, Z., Raison, M., Tejani, A., Chilamkurthy, S.,
Steiner, B., Fang, L., \ldots{} Chintala, S. (2019). {PyTorch: An
Imperative Style, High-Performance Deep Learning Library}. \emph{arXiv
e-Prints}, arXiv:1912.01703.
\url{https://doi.org/10.48550/arXiv.1912.01703}

\bibitem[\citeproctext]{ref-pinte18}
Pinte, C., Price, D. J., Ménard, F., Duchêne, G., Dent, W. R. F., Hill,
T., de Gregorio-Monsalvo, I., Hales, A., \& Mentiplay, D. (2018).
{Kinematic Evidence for an Embedded Protoplanet in a Circumstellar
Disk}. \emph{The Astrophysical Journal Letters}, \emph{860}(1), L13.
\url{https://doi.org/10.3847/2041-8213/aac6dc}

\bibitem[\citeproctext]{ref-smirnov11a}
Smirnov, O. M. (2011). {Revisiting the radio interferometer measurement
equation. I. A full-sky Jones formalism}. \emph{527}, A106.
\url{https://doi.org/10.1051/0004-6361/201016082}

\bibitem[\citeproctext]{ref-tazzari18}
Tazzari, M., Beaujean, F., \& Testi, L. (2018). {GALARIO: a GPU
accelerated library for analysing radio interferometer observations}.
\emph{476}, 4527--4542. \url{https://doi.org/10.1093/mnras/sty409}

\bibitem[\citeproctext]{ref-thompson17}
Thompson, A. R., Moran, J. M., \& Swenson, Jr., George W. (2017).
\emph{{Interferometry and Synthesis in Radio Astronomy, 3rd Edition}}.
\url{https://doi.org/10.1007/978-3-319-44431-4}

\bibitem[\citeproctext]{ref-wang23}
Wang, R., Chen, Z., Luo, Q., \& Wang, F. (2023). {A Conditional
Denoising Diffusion Probabilistic Model for Radio Interferometric Image
Reconstruction}. \emph{arXiv e-Prints}, arXiv:2305.09121.
\url{https://doi.org/10.48550/arXiv.2305.09121}

\bibitem[\citeproctext]{ref-wootten09}
Wootten, A., \& Thompson, A. R. (2009). {The Atacama Large
Millimeter/Submillimeter Array}. \emph{IEEE Proceedings}, \emph{97}(8),
1463--1471. \url{https://doi.org/10.1109/JPROC.2009.2020572}

\bibitem[\citeproctext]{ref-zawadzki23}
Zawadzki, B., Czekala, I., Loomis, R. A., Quinn, T., Grzybowski, H.,
Frazier, R. C., Jennings, J., Nizam, K. M., \& Jian, Y. (2023).
{Regularized Maximum Likelihood Image Synthesis and Validation for ALMA
Continuum Observations of Protoplanetary Disks}. \emph{Publications of
the Astronomical Society of the Pacific}, \emph{135}(1048), 064503.
\url{https://doi.org/10.1088/1538-3873/acdf84}

\end{CSLReferences}
\end{document}